\documentclass[aps,pra,twocolumn,superscriptaddress,showpacs,reprint]{revtex4-1}
\usepackage{graphicx}
\usepackage{amsfonts}
\usepackage{amssymb}
\usepackage{array}
\usepackage{amsmath}
\usepackage{color}
\usepackage{verbatim}
\usepackage{slashed}
\usepackage{subfigure}
\usepackage{subfloat}
\usepackage{epstopdf}

\makeatletter
\renewcommand{\maketag@@@}[1]{\hbox{\m@th\normalsize\normalfont#1}}%
\newcommand*{\QEDA}{\hfill\ensuremath{\blacksquare}}%
\makeatother

\begin{document}
\title{Additivity relations in quantum correlations}
\author{Seungho Yang}
\affiliation{Center for Macroscopic Quantum Control,
Department of Physics and Astronomy,
Seoul National University, Seoul, 151-742, Korea}

\author{Hyunseok Jeong}
\affiliation{Center for Macroscopic Quantum Control,
Department of Physics and Astronomy,
Seoul National University, Seoul, 151-742, Korea}

\author{Wonmin Son}
\affiliation{Department of Physics, Sogang University, Seoul, 121-742, Korea}

\date{\today}

\begin{abstract}
Does the sum of correlations in subsystems constitute the correlation in total system? Such a concept can be expressed by an {\it additivity relationship} of correlations. From {\it strong subadditivity} condition of von Neumann entropy, four different additivity relations in {\it total correlation} are derived and quantified. Based upon the classification of the additivity in total correlation, we identify the corresponding additive relationships in {\it entanglement}. It is also discussed that similar relationships are satisfied for quantum discord of pure states, but it is not always true for mixed states.
\end{abstract}

\pacs{03.65.Ud, 03.67.Mn}
\maketitle

\section{introduction}
Quantifying correlations in quantum systems is an important issue in quantum information theory. From an hierarchical picture, measures of correlations can be classified into three parts: total correlations, quantum correlations, classical correlations \cite{Groisman05}. Roughly specking, {\it total correlation} encapsulates all the correlations in a composite system, including quantum and classical correlations. Contrarily, {\it entanglement} is a well-known measure of quantum correlation which does not take into account the classical counterpart \cite{Horodecki09}. It has been identified that entanglement is a necessary condition for providing exponential speed-up in quantum computation \cite{Jozsa03} and is taken as a useful resource for various quantum information processing. Recently, another class of quantum correlation, {\it quantum discord}, is proposed \cite{Ollivier01,Modi12}. It has been argued that discord characterizes quantum correlation better in terms of computational advantage and the operational meanings of discord have recently begun to be understood \cite{Dakic12,Gu12,Madhok12,Datta08}.

From the spirit of multipartite correlation which quantifies overall amount of correlation, it is natural to think about the relationships between the multipartite correlation in total system and the bipartite correlations of subsystems. A basic example is that, for a tripartite system A-B-C, tripartite correlation is required to be equal to or greater than bipartite correlation in any subsystem. Also, a measures of tripartite correlations should be reduced to a measures of bipartite correlations in a subsystem when a party has completely no correlation with the other parties. For the case of total correlation, such relationships are well-defined by additivity relations. We are questioning here whether such relations are generally satisfied for different measures of {\it quantum correlations}, e.g., entanglement and discord.

Formal definition of total correlation is given as an optimal distance of entropy between a given state and a state without any correlation, {\it product state}. A multipartite state is called product state if the state of a $n$-party quantum system $\mathcal{A}_{1}\cdot\cdot\cdot\mathcal{A}_{n}$ is written as a product, $\pi =\pi_{1}\otimes\pi_{2}\otimes...\otimes\pi_{n}$ where $\pi_i$ is a positive operator in the Hilbert space $\mathcal{A}_i$. The total information can be represented as
\begin{equation}
\label{eq:total}
\mathcal{T}(\rho)=\min_{\substack{\sigma}\in \pi} S(\rho \vert\vert\sigma)
\end{equation}
where $ S(\rho \vert\vert\sigma)=\mbox{Tr}\left[ \rho\log\rho-\rho\log\sigma\right]$ is the relative entropy and ``$\min$'' denotes the minimization of $S(\rho \vert\vert\sigma)$ over $\sigma$ within a product state $\pi$. After a simple algebra, it can be proved that the closest product state to a state $\rho\in\mathcal{A}_{1}\cdot\cdot\cdot\mathcal{A}_{n}$ is the product of reduced states, $\rho_{1}\otimes\rho_{2}\otimes...\otimes\rho_{n}$ \cite{Modi10} where $\rho_i=\mbox{Tr}_{i|All} [\rho]$ and $\mbox{Tr}_{i|All}$ means trace out of all the state except the state in the $i$th Hilbert space. Therefore, for an arbitrary multipartite state, it can be written in a different way as $\mathcal{T}(\rho)= S(\rho\vert\vert\rho_{1}\otimes\rho_{2}\otimes...\otimes\rho_{n})=\sum_i S(\rho_i)-S(\rho)$ where $S(\rho)=-\mbox{Tr}[\rho \log\rho]$ is von Neumann entropy \cite{Modi10}. The quantity is sometimes called as total mutual information since the total information becomes mutual information when the system is bipartite, as $I(\rho_{A,B})=S(\rho_{A})+S(\rho_{B})-S(\rho_{A,B})$ defined in a bipartite system $A-B$ \cite{Adami97,Li07,Groisman05}. In terms of the quantum relative entropy, $I(\rho_{A,B})\equiv S(\rho_{A,B}\vert\vert\rho_{A}\otimes\rho_{B})$.

A physical state $ \mathcal{S}$ of $n$-party quantum system $\mathcal{A}_{1}\cdot\cdot\cdot\mathcal{A}_{n}$ is called separable if the state can be written as $\mathcal{S}=\sum_{\substack{i}}p_{i}s_{1}^{i}\otimes...\otimes s_{n}^{i}$ where $s_{m}^{i}$ is a state in $\mathcal{A}_{m}$. By definition, the state is considered to have no entanglement and it is a type of multipartite state which can be prepared in a classical methods, i.e., local operation and classical communication. Under the circumstance, a measure of entanglement can be characterized using a distance measure between a given state and the closest separable state, called the relative entropy of entanglement \cite{Vedral02,Vedral97,Vedral98}. The relative entropy of entanglement is defined as
\begin{equation}
\label{eq:relative}
\mathcal{E}(\rho)=\min_{\substack{\sigma}\in  \mathcal{S}}S(\rho \vert\vert\sigma)
\end{equation}
where $\mathcal{S}$ is the set of separable states. The measure of entanglement is naturally applied to multipartite systems as the criterion of separable states has been well-defined for a multipartite system.

In addition, the notion of quantum correlation can be greatly extended once the notion of state distance from classically correlated states instead of separable states is considered \cite{GROISMAN07,SAITOH08,Bravyi03,Luo08,Gharibian11,Modi10}. A physical state $\rho$ of a $n$-party quantum system $\mathcal{A}_{1}\cdot\cdot\cdot\mathcal{A}_{n}$ is called classically correlated if the state can be written as $\mathcal{C}=\sum_{\substack{\vec{k}}}p_{\vec{k}}\vert\vec{k}\rangle\langle\vec{k}\vert$ where $\vert\vec{k}\rangle=\vert k_1 \rangle\otimes\vert k_2 \rangle \cdot\cdot\cdot \otimes \vert k_n \rangle$ and $\vert k_l \rangle$ is a local basis at $l$-th site. The states are considered to have no entanglement since they are convex sum of product states. The relative entropy of discord \cite{Modi12,GROISMAN07,SAITOH08,Bravyi03,Luo08,Gharibian11,Modi10} is defined to be
\begin{equation}
\label{eq:discord}
\mathcal{D}(\rho)=\min_{\substack{\sigma}\in  \mathcal{C}}S(\rho \vert\vert\sigma)
\end{equation}
where $\mathcal{C}$ is the set of classically correlated states. Due to its definition, the measure of discord is also naturally applied to multipartite systems.

As it was seen from their definition, the total mutual information, the relative entropy of entanglement, and the relative entropy of discord are multipartite measures of correlations in a unified view of correlations. Those measures have the clear relationship $\mathcal{T}\geq\mathcal{D}\geq\mathcal{E}$ because $\pi \subset \mathcal {C} \subset \mathcal {S}$, and they also enable us to compare the three criteria of correlation, product states, classically correlated states, and separable states.


In the following section, we discuss about the additivity relation in the total correlation, entanglement and discord. We derive four different additivity relationships using the total correlation (\ref{eq:total}) and shows that whether similar type of correlation can be derived for the case of entanglement and discord. In the section III, it is shown that the additivity relationship is satisfied by a specific class of states. As specific examples, we will deal the generalized GHZ-states, a variant of the generalized GHZ-states, the generalized W-states, and a few mixed states. In the section IV, we summarize our result and presents the conclusion on the additivity relationship in quantum correlations.

\section{Additive relations in multipartite correlation}
In this section, we discuss about the additivity relation of multipartite correlations. At first, we discuss that how a multipartite correlation in general can be related with sets of bipartite correlations. We will show that the relationship is led to three different types of inequalities. It is our main question whether such inequalities are satisfied for the case of quantum correlations so that we will study the relationships between multipartite and bipartite quantum correlations using entanglement as well as quantum discord.

\subsection{Bounds on the total correlation}
The total information counts all the correlation existing in a multipartite system. The structure of multipartite correlation is quite complicated in general although there are rules by which they can be decomposed into the correlations of subsystems. In fact, the total mutual information of a tripartite A-B-C system can in principle be decomposed by two different types of bipartite correlations as such
\begin{align}
\label{total equality}
\mathcal{T}(A:B:C)=\mathcal{T}(A:B)+\mathcal{T}(AB:C)
\end{align}
and the equality holds for any permutation of A, B, and C. The equation shows how the tripartite correlation is quantified in terms of bipartite correlations. If we consider only the subsystem A-B among the total system, there is only correlation between A and B. In addition, if a system C is joined to the subsystem A-B, it is natural to understand that the total amount of correlation is increased and the amount corresponds to the correlation existing between the subsystem A-B together with the correlation between A-B and the additional system C. The additional correlation is taken by the term $\mathcal{T}(AB:C)$ while it is still characterized by the bipartite correlation between two different subsystems, A-B and C.

The relation between the multipartite correlation and the bipartite correlations has been well summarized in the famous strong subadditivity relation of Shannon entropy $$H(AB) +H(BC) \ge H(B)+H(ABC)$$ where $H(p)=-\sum_i p_i\log p_i$ \cite{footnote1}. The inequality is easy to prove in general, while the quantum version of the strong subadditivity with von Neumann entropy is difficult. The strong subadditivity of von Neumann entropy $S(\rho)=-\mbox{Tr}[\rho \log\rho]$ \cite{Lieb73} refers to the following inequality for tripartite quantum state $\rho_{ABC}$.
\begin{align*}
S(AB)+S(BC)\geq S(ABC)+S(B)
\end{align*}
where $\rho_{AB}=\operatorname{tr}_{C}(\rho_{ABC})$, $\rho_{BC}=\operatorname{tr}_{A}(\rho_{ABC})$ and $\rho_{B}=\operatorname{tr}_{A,B}(\rho_{ABC})$. The strong subadditivity is widely used in quantum information theory \cite{Nielsen00}. The proof of the inequality and its operational meaning can be found in \cite{Lieb73,Wehrl78,Schumacher96}.

\begin{figure}[t]
\centering
\subfigure[][~Decomposition of tripartite correlation.]{\includegraphics[width=0.35\textwidth]{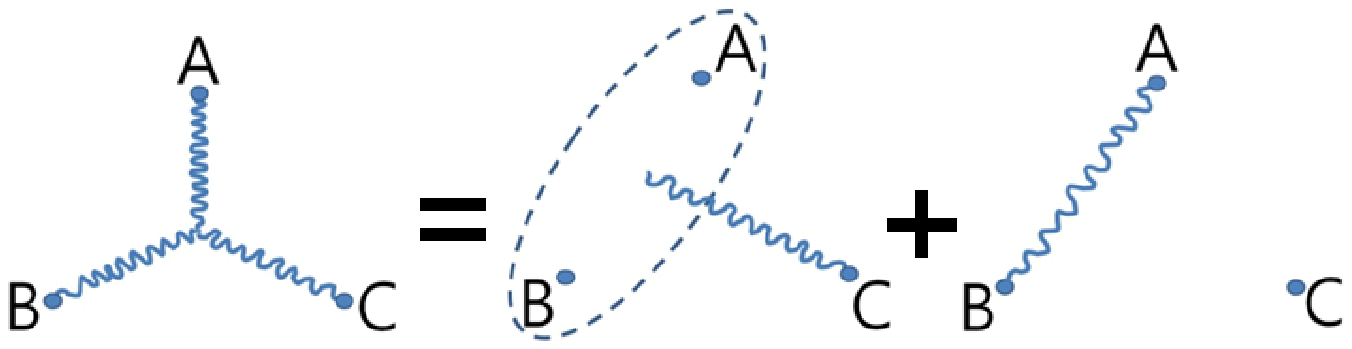}\label{fig_a}}
\subfigure[][~Lower bound for bipartition.]{\includegraphics[width=0.35\textwidth]{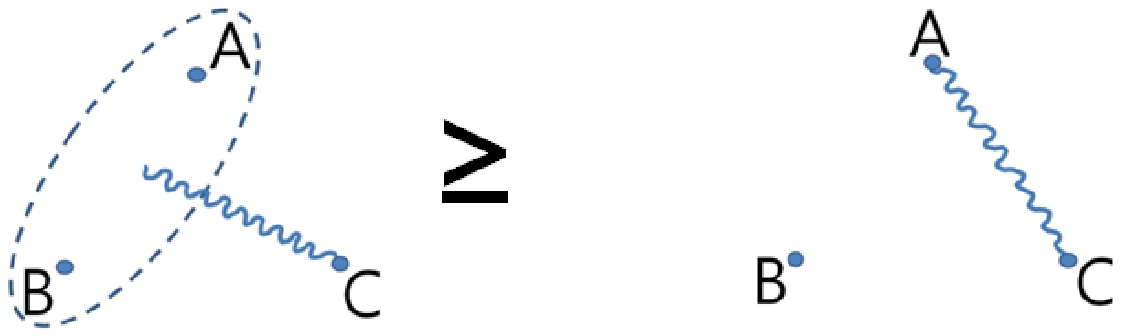}\label{fig_b}}
\subfigure[][~Lower bound in tripartite correlation.]{\includegraphics[width=0.35\textwidth]{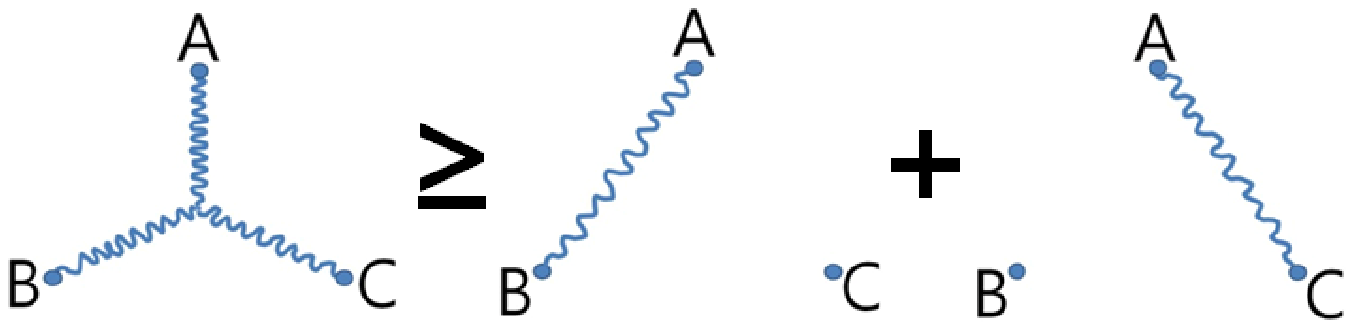}\label{fig_c}}
\subfigure[][~Upper bound in tripartite correlation.]{\includegraphics[width=0.35\textwidth]{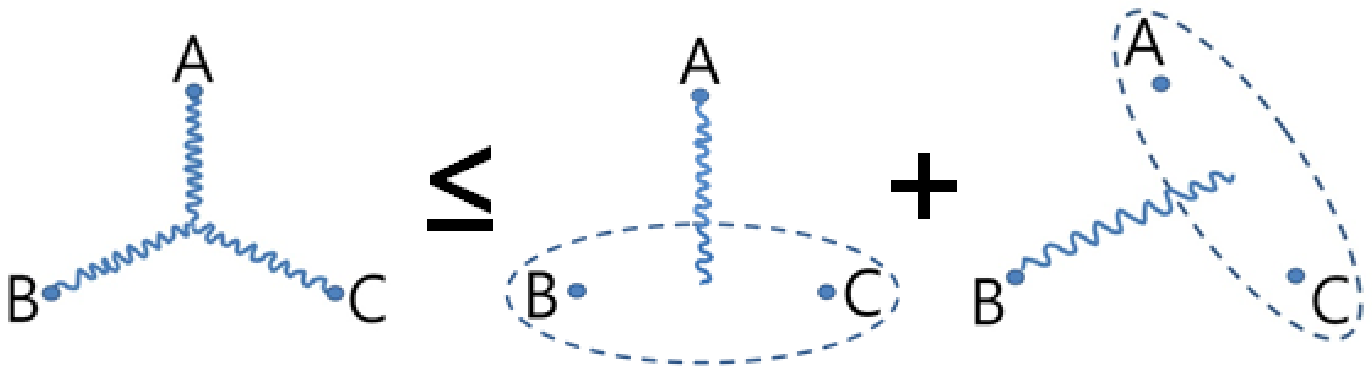}\label{fig_d}}
\caption{Conceptual diagrams for the relations between the correlations in a three-party system A-B-C. Sub-figures (a) to (d) correspond to Eqs.~\eqref{total equality} to \eqref{total bound 3} respectively.}
\label{conceptual}
\end{figure}

Using the strong subadditivity of von Neumann entropy, the bounds on the total mutual information are obtained as
\begin{eqnarray}
\mathcal{T}(AB:C)&\geq&\mathcal{T}(A:C),\label{total bound 1}\\
\mathcal{T}(A:B:C)&\geq& \mathcal{T}(A:B)+\mathcal{T}(A:C), \label{total bound 2}\\
\mathcal{T}(A:B:C)&\leq&\mathcal{T}(BC:A)+\mathcal{T}(AC:B). \label{total bound 3}
\end{eqnarray}
From the definition, it is straightforward to show that the three inequalities above are equivalent to the strong subadditivity of von Neumann entropy. The strong subadditivity frequently has been represented in the form of Eq.~\eqref{total bound 1} \cite{Nielsen00}, but it can also be represented in the forms of Eqs.~\eqref{total bound 2} and \eqref{total bound 3} by using the tripartite total mutual information. We can give intuitive interpretations of the three inequalities. The first inequality states that the correlation between A-B and C contains the correlation between A and C. The correlation between A and B is missing in the right-hand side. The second inequality is also intuitively clear because the left-hand side of the inequality counts all the correlations among A, B, and C, so it is greater than or equal to the correlations between A-B and A-C only. In that case, it can be said that the correlation between B and C has not been counted. In the third inequality, both two terms of the right-hand side contain the correlation between A and C (i.e., double counting),  so it is less than or equal to the tripartite correlation.
Those relations between the correlations in a tripartite system A-B-C are represented in the conceptual diagrams in Fig. \ref{conceptual}.

\subsection{Bounds on the relative entropy of entanglement}
The total information covers all the correlations in both a classical part and a quantum part. One may wonder that there exist the consistent inequalities as (\ref{total equality}), (\ref{total bound 1}), (\ref{total bound 2}) and (\ref{total bound 3}) for the measures of quantum correlations. A natural candidate for quantum correlation is quantum entanglement. There are several measures of entanglement which have been studied extensively so far \cite{Horodecki09}. However, our question requires a measure which can be naturally generalized to multipartite systems and which is consistent to the quantum part of total information in Eq.~\eqref{eq:total}. A proper measure in that purpose is the relative entropy of entanglement which is defined in Eq.~(\ref{eq:relative}). The relative entropy of entanglement is always less than or equal to the total information and is naturally generalized to multipartite systems.

For the relative entropy of entanglement, there exists an inequality which corresponds to Eq.~\eqref{total equality} of the total information. The authors of \cite{Plenio01} first showed that the following inequality holds for any tripartite pure system A-B-C.
\begin{align}\label{entanglement equality}
\mathcal{E}(A:B:C) \geq& \mathcal{E}(A:B)+\mathcal{E}(AB:C).
\end{align}
They derived it by using the inequality $S(\sigma_{ABC}\vert\vert\rho_{ABC})-S(\sigma_{AB}\vert\vert\rho_{AB})\geq S(\sigma_{AB})-S(\sigma_{ABC})$ \cite{Plenio00}, which holds for any state $\rho_{ABC}$ and for any tri-separable state $\sigma_{ABC}$. Unlike the equation \eqref{total equality} of the total mutual information, Eq.~\eqref{entanglement equality} is an inequality. It means that the tripartite entanglement contains entanglement between A, B and AB, C but they are not exactly same.

Also, the following inequality holds for any tripartite system A-B-C:
\begin{align}\label{entanglement bound1}
\mathcal{E}(AB:C)\geq \mathcal{E}(A:C).
\end{align}
It can be derived directly from the monotonicity of the relative entropy under partial trace, $S(\rho^{A}\vert\vert \sigma^{A})\leq S(\rho^{AB}\vert\vert\sigma^{AB})$ \cite{Nielsen00}.

Combining Eqs.~\eqref{entanglement equality} and \eqref{entanglement bound1} gives the following inequalities which hold for any tripartite pure system A-B-C:
\begin{equation}
\mathcal{E}(A:B:C)\geq \mathcal{E}(A:B)+\mathcal{E}(A:C).
\label{entanglement bound2}
\end{equation}
At the same time, a symmetrized version of the inequality can be found as
$\mathcal{E}(A:B:C)\geq \frac{2}{3}\big[\mathcal{E}(A:B)+\mathcal{E}(B:C)+\mathcal{E}(A:C)\big]$ by taking the average of Eq.~\eqref{entanglement bound2} over all permutations of A, B, and C.

The inequality \eqref{entanglement bound2} corresponds to Eq.~\eqref{total bound 2} of the total mutual information. It is a weaker inequality than the monogamy inequality of entanglement $\mathcal{E}(A:BC)\geq \mathcal{E}(A:B)+\mathcal{E}(A:C)$, which is not known if it holds for the entanglement measure. It is not an unique property of quantum correlations. We can see the analogy in the classical correlations between classical random variables. For three classical random variables X, Y and Z, $H(X:Y:Z)\geq H(X:Y)+H(X:Z)$ where $H(X:Y:Z)=H(X)+H(Y)+H(Z)-H(X,Y,Z)$ is a multivariate mutual information.

There is also the same form of upper bound for the regularized relative entropy of entanglement $\mathcal{E}^{\infty}(\rho)=\lim_{n\rightarrow \infty } \frac{1}{n} \mathcal{E}(\rho^{\otimes n})$,
\begin{align*}
\mathcal{E}^{\infty}(A:B:C)\leq \mathcal{E}^{\infty}(BC:A)+\mathcal{E}^{\infty}(AC:B),
\end{align*}
which holds for any tripartite pure system. The inequality is also proven in \cite{Plenio01}, and it follows from the fact the $\mathcal{E}$ is an entanglement monotone.

Therefore, it is possible to conclude that similar kinds of additivity relationship with total correlation are satisfied for the case of entanglement. Although it is found that they do not coincide sharply, they asymptotically behave in the same way.

\subsection{Bounds on the relative entropy of discord}
 Now we consider the quantum discord in Eq.~(\ref{eq:discord}) as another candidate of quantum correlations. It is well-known that the criterion of non-zero quantum discord is not identical to the criterion of non-zero entanglement. Unlike entanglement, quantum discord is defined to be non-zero even when a state is separable. The difference makes it non-trivial to consider whether there exist the consistent additivity relations as Eqs.~(\ref{total equality}) to (\ref{total bound 3}) for quantum discord in multipartite systems. We take the relative entropy of discord in Eq.~(\ref{eq:discord}) as quantum discord measure. The relative entropy of discord differs from the relative entropy of entanglement in the set of states whose minimization should be taken. It enable us to genuinely compare the difference in the separability criterion and the classicality criterion.

 First, we show that there is the consistent lower bound on the tripartite relative entropy of discord. The minimization in $\mathcal{D}(\rho)=\min_{\substack{\sigma}\in \mathcal{C}}S(\rho \vert\vert\sigma)$ can be reduced to a minimization over set of local bases $\{\vert \vec{k} \rangle\}$ \cite{Modi10}
\begin{align}\label{quantumness 2}
\mathcal{D}(\rho)=\min_{\{\vert\vec{k}\rangle\}}\Lambda_{\rho}(\{\vert\vec{k}\rangle\})-S(\rho)
\end{align}
where $\Lambda_{\rho}(\{\vert\vec{k}\rangle\})=-\sum_{\vec{k}}\langle \vec{k} \vert \rho \vert \vec{k}\rangle \log \langle \vec{k} \vert \rho \vert \vec{k}\rangle$. For the matter of convenience, we write $A:B:C$ instead of $\rho_{ABC}$ from now on. A:B:C denotes a tripartition of a system A-B-C.

\emph{Theorem 1.}
The following inequality holds for any tripartite pure system A-B-C.
\begin{align}\label{discord equality}
\mathcal{D}(A:B:C)&\geq \mathcal{D}(A:B)+\mathcal{D}(AB:C).
\end{align}

\emph{proof} ---  For any pure state, bipartite discord is equivalent to the entropy of entanglement, so $\mathcal{D}(AB:C)=S(AB)=S(C)$. Then the inequality is equivalent to $\min\Lambda_{A:B:C}(\{\vert\vec{k}\rangle\}) \geq \min\Lambda_{A:B}(\{\vert\vec{k}\rangle\})$. Let $\{\vert\vec{k}^{*}\rangle\}$ be a basis which minimizes $\Lambda_{A:B:C}(\{\vert\vec{k}\rangle\})$, and consider the set of local measurement operators $\{ \vert \vec{k}^{*} \rangle \langle \vec {k}^{*} \vert\}$. The outcomes of the local measurements can be treated as three classical random variables, which are denoted X, Y and Z. They correspond to the outcomes of the local measurements on A, B and C respectively. For any classical random variables X, Y and Z, Shannon entropies satisfy $H(X,Y,Z)\geq H(X,Y)$. Using this inequality, $\min\Lambda_{A:B:C}(\{\vert\vec{k}\rangle\})=H(X,Y,Z)\geq H(X,Y)\geq \min\Lambda_{A:B}(\{\vert\vec{k}\rangle\})$, so the proof is completed. \QEDA

 We also found the analogous inequality for discord as the inequality \eqref{total bound 1} of the total mutual information and the inequality \eqref{entanglement bound1} of the relative entropy of entanglement.

\emph{Theorem 2.} The following inequality holds for any tripartite pure system A-B-C.
\begin{align}
\label{theorem 2}
\mathcal{D}(AB:C)&\geq \frac{1}{2} \Big[\mathcal{D}(A:C)+\mathcal{D}(B:C)\Big],
\end{align}
and its proof can be given in the below.

\emph{proof} --- The inequality \eqref{theorem 2} can be easily proven from just $\mathcal{T}\geq\mathcal{D}$. Given a pure system A-B-C, the bipartite discord $\mathcal{D}(AB:C)$ is equivalent to the entropy of entanglement $S(C)$, and $S(AC)=S(B)$ and $S(BC)=S(A)$. Therefore, $S(C)=\frac{1}{2}[\mathcal{T}(A:C)+\mathcal{T}(B:C)]\geq\frac{1}{2}[\mathcal{D}(A:C)+\mathcal{D}(B:C)]$ from $\mathcal{T}\geq\mathcal{D}$. \QEDA

\emph{Corollary.} Consider a tripartite pure system A-B-C whose parties are ordered so $\mathcal{D}(A:B)\geq\mathcal{D}(B:C)\geq\mathcal{D}(A:C)$. Then the following inequalities hold.
\begin{align}
\mathcal{D}(AB:C)&\geq \mathcal{D}(A:C), \label{c1} \\
\mathcal{D}(A:B:C)&\geq \mathcal{D}(A:B)+\mathcal{D}(A:C), \label{c2} \\
\mathcal{D}(BC:A)+\mathcal{D}(AC:B)&\geq \mathcal{D}(A:B)+\mathcal{D}(A:C). \label{c3}
\end{align}

\emph{proof} --- They are directly followed by Eqs.~\eqref{discord equality}, \eqref{theorem 2}, and the assumption $D(A:B) \ge D(B:C)\ge D(A:C)$. On the other hand, by taking the average of \eqref{c2} over all permutations of three parties, one can obtain $\mathcal{D}(A:B:C)\geq \frac{2}{3}[\mathcal{D}(A:B)+\mathcal{D}(B:C)+\mathcal{D}(A:C)]$.

\begin{figure}
\centering
\subfigure{\includegraphics[width=0.31\textwidth]{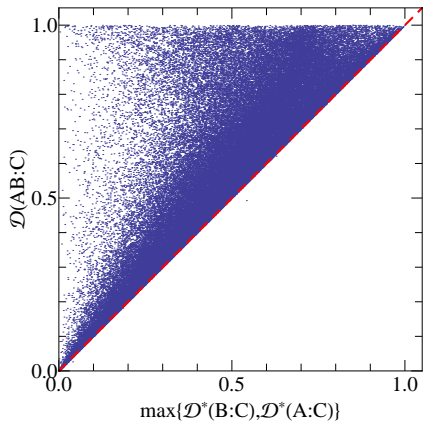}}
\caption{(Color Online) Numerical tests of the inequality \eqref{discord bound1} for $10^{6}$ random samples of three-qubit pure states. No sample was found violating. $\mathcal{D}^{*}$ is approximated discord obtained from numerical minimization. Rigorously speaking, $\mathcal{D}^{*}$ is equal to or greater than $\mathcal{D}$.}\label{p1}
\end{figure}

 In order that Eqs.~\eqref{c1}, \eqref{c2}, and \eqref{c3} are satisfied for any permutation of A, B, and C, one needs to show that
\begin{align}
\mathcal{D}(AB:C)\geq \max \{\mathcal{D}(B:C),\mathcal{D}(A:C)\} \label{discord bound1}
\end{align}
 for any tripartite pure system A-B-C. Its analytical proof is not found. However, we conjecture with numerical evidence that it holds for any three-qubit pure state. For pure states, $\mathcal{D}(AB:C)=S(C)$ and $\mathcal{D}(A:C)=\min \Lambda_{A:C}(\vert \vec{k}\rangle)-S(B)$. The main task of the numerical computations is the minimization of $\Lambda_{A:C}(\vert \vec{k}\rangle)$ over all local bases. In general, there are several local minima in $\Lambda(\vert \vec{k}\rangle)$, so the results of the numerical minimizations can not guarantee that it always finds the global minimum; it just provides an upper bound. Fortunately, however, the part consisting the minimization is on the right-hand side of the inequality \eqref{discord bound1}, so we can provide numerical evidence for the inequality by testing numerous samples. We compare the both sides in Eq.~\eqref{discord bound1} for $10^{6}$ random samples of pure three-qubit states. No quantum state has been found violating Eq.~\eqref{discord bound1}. The result is given in Fig. \ref{p1}. The random samples are generated from the parameterization \cite{Acin00} of three-qubit states $\vert\psi\rangle=\lambda_{0}\vert000\rangle+e^{i\phi}\lambda_{1}\vert100\rangle+\lambda_{2}\vert101\rangle+\lambda_{3}\vert110\rangle+\lambda_{4}\vert111\rangle$. To minimize Eq.~\eqref{quantumness 2}, we parameterized an arbitrary local basis $\{\vert 0^{\prime}\rangle,\vert1^{\prime}\rangle\}$ as
\begin{equation}\label{parameterization}
\begin{split}
\vert0\rangle &= \sqrt{1-t^{2}}\vert0^{\prime}\rangle+t\vert1^{\prime}\rangle \\
\vert1\rangle &= e^{i\phi}(t\vert0^{\prime}\rangle-\sqrt{1-t^{2}}\vert1^{\prime}\rangle),
\end{split}
\end{equation}
where $0\leq e^{i\phi}\leq 2\pi$ and $0\leq t\leq1$. In fact, 4 real parameters are required to cover all bases, but it can be reduced to 2 as we only consider $\Lambda_{\rho}(\{\vert\vec{k}\rangle\})$. The number of parameters is two per each qubit, so computing $\min\Lambda_{A:C}(\{\vert\vec{k}\rangle\})$ requires numerical minimization over 4 parameters.

The inequality \eqref{c1} is suggestive of the monogamy inequality \cite{Prabhu11,Giorgi11,Streltsov11}. We note that they look similar, but are different. The monogamy of entanglement is an unique property of quantum states, and it is not found in correlations of classical variables. In contrast, Eq.~\eqref{c3} also holds for the mutual information of classical variables. The inequality \eqref{c3} is studied for the global discord in Ref.~\cite{Braga12}, and it is shown that Eq.~\eqref{c1} is a sufficient condition for Eq.~\eqref{c3}.

For general mixed states, we could not find any example violating Eqs.~\eqref{c1}, \eqref{c2}, or \eqref{c3}. However, we note that they can be violated by the other permutations of A, B, and C. Bipartite discord of some mixed states can increase after discarding a subsystem, so Eq.~\eqref{discord bound1} dose not hold for general mixed states. Consider the state $\rho=p\vert000\rangle\langle000\vert + (1-p)\vert1\mathcal{+}1\rangle\langle1\mathcal{+}1\vert$ where $\vert + \rangle = 1/\sqrt{2}(\vert 0\rangle +\vert 1 \rangle)$. For the state, $\mathcal{D}(BC:A)=0$ and $\mathcal{D}(A:C)>0$, so $\mathcal{D}(BC:A)\ngeq\mathcal{D}(A:C)$ for any measure of discord. In our measure of discord, $\mathcal{D}(BC:A)=0$ and $\mathcal{D}(A:C)=\mathcal{D}(A:B:C)=\min\{p,(1-p)\}$. We leave the details of the calculation to Appendix. In contrast, we mentioned that the relative entropy of entanglement always satisfies $\mathcal{E}(AB:C)\geq\mathcal{E}(B:C)$. The difference of entanglement and discord occurs from the criteria of classically correlated states and separable states. If $\rho_{ABC}$ is separable in the bipartition AB and C, then $\rho_{AC}$ and $\rho_{BC}$ are also separable. However, $\rho_{AC}$ and $\rho_{BC}$ can be not classically correlated although $\rho_{ABC}$ is classically correlated in the bipartition AB and C.

On the other hand, we leave a unanswered question of whether the inequality $\mathcal{D}(A:B:C)\leq\mathcal{D}(BC:A)+\mathcal{D}(AC:B)$ holds for tripartite systems A-B-C whose parties are ordered so $\mathcal{D}(A:B)\geq\mathcal{D}(B:C)\geq\mathcal{D}(A:C)$.  There is an example violating it for the other permutations of A, B, and C. For the state $\rho=p\vert000\rangle\langle000\vert + (1-p)\vert1\mathcal{+}1\rangle\langle1\mathcal{+}1\vert$, $\mathcal{D}(AB:C)+\mathcal{D}(BC:A)=0$, but $\mathcal{D}(A:B:C)=\min\{p,(1-p)\}$.

We have discussed about the consistent additivity relations for the multipartite measures of total correlations and quantum correlations. It would also be worth to consider the relations for multipartite measures of classical correlations. In terms of the quantum relative entropy, a measure of classical correlations is defined \cite{Modi10} as
\begin{align}
\mathcal{C}(\rho)&=\min_{\sigma\in\pi}S(\mathcal{\chi}_{\rho}\vert\vert\sigma).\label{classical correlation}
\end{align}
Here, $\mathcal{\chi}_{\rho}$ is the closest classically correlated state to $\rho$. It is not known whether the measure of classical correlations also has the consistent additivity relations as the measures of total correlations and quantum correlations. The part is left as an open question. On the other hand, it is evident that for the classically correlated states, $\rho\in \mathcal{C}$, those measures also follow the relations \eqref{total equality}, \eqref{total bound 1}, \eqref{total bound 2}, and \eqref{total bound 3} of the total mutual information because the classical correlation is just the total correlation.

\section{Specific examples of additivity relations}
In the preceding section, we have shown that the additivity relations from Eq.~\eqref{total equality} to Eq.~\eqref{total bound 3} are also satisfied by entanglement and discord [for discord, Eq.~\eqref{total bound 3} was left unsolved, and only Eq.~\eqref{c3} is proved] with help of numerical analysis, especially when the total system is in a three-qubit pure state.
We note that the equality \eqref{total equality} for the total mutual information
becomes an inequality (i.e., ``$=$'' is replaced with ``$\geq$'') when applied to entanglement and discord.
In subsection A, we now present several examples of three-qubit pure states
in which all of multipartite entanglement, discord and classical correlations satisfy all the
the additivity relations from \eqref{total bound 1} to \eqref{total bound 3} as well as
the inequality \eqref{discord equality}. In the following section, we have tried to identify the validity of additivity relationship when the total system is in a mixed state. It was possible to find specific cases where the additivity relationship is not satisfied.

\subsection{Additivity relations of discord in pure states}
We first consider a GHZ state in the form of
\begin{align*}
\vert GHZ\rangle=\frac{1}{\sqrt{2}}(\vert000\rangle+\vert111\rangle).
\end{align*}
Both the tripartite entanglement and discord of the GHZ state
are equal to 1, while both of the bipartite entanglement and discord are zero for any bipartite subsystem. Accordingly,
if we let $\mathcal{Q}$ represent either
$\mathcal{E}$ or $\mathcal{D}$,
$\mathcal{Q}(A:B:C)=\mathcal{Q}(AB:C)=1$ and $\mathcal{Q}(A:B)=0$ for any permutation of $A$, $B$, and $C$, so that all the additivity relations from \eqref{total equality} to
\eqref{total bound 3} are satisfied.
In the case of a more general form, $\vert\psi\rangle=\alpha\vert000\rangle+\beta\vert111\rangle$,
it is straightforward to obtain $\mathcal{Q}(A:B:C)=\mathcal{Q}(AB:C)=-\vert\alpha\vert^{2}\log\vert\alpha\vert^{2}-\vert\beta\vert^{2}\log \vert\beta\vert^{2}$ and $\mathcal{Q}(A:B)=0$.
All the additivity relations are then satisfied for any permutation. In addition, we can calculate classical correlations $\mathcal{C}$ in Eq.~\eqref{classical correlation}. Using the fact that the closest product state to a given state is the product of its reduced states \cite{Modi10}, we find $\mathcal{C}(A:B:C)/2=\mathcal{C}(AB:C)=\mathcal{C}(A:B)=-\vert\alpha\vert^{2}\log\vert\alpha\vert^{2}-\vert\beta\vert^{2}\log \vert\beta\vert^{2}$. Again, all the additivity relations are met.

\begin{figure}
\centering
\subfigure[][]{\includegraphics[width=0.23\textwidth]{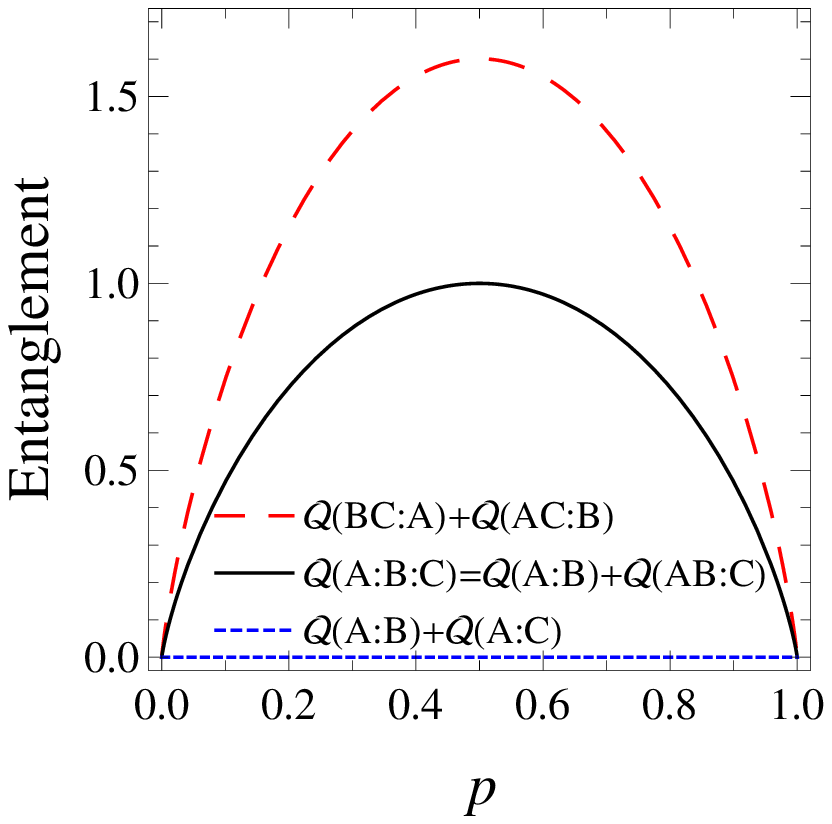}\label{S-a}}
\subfigure[][]{\includegraphics[width=0.23\textwidth]{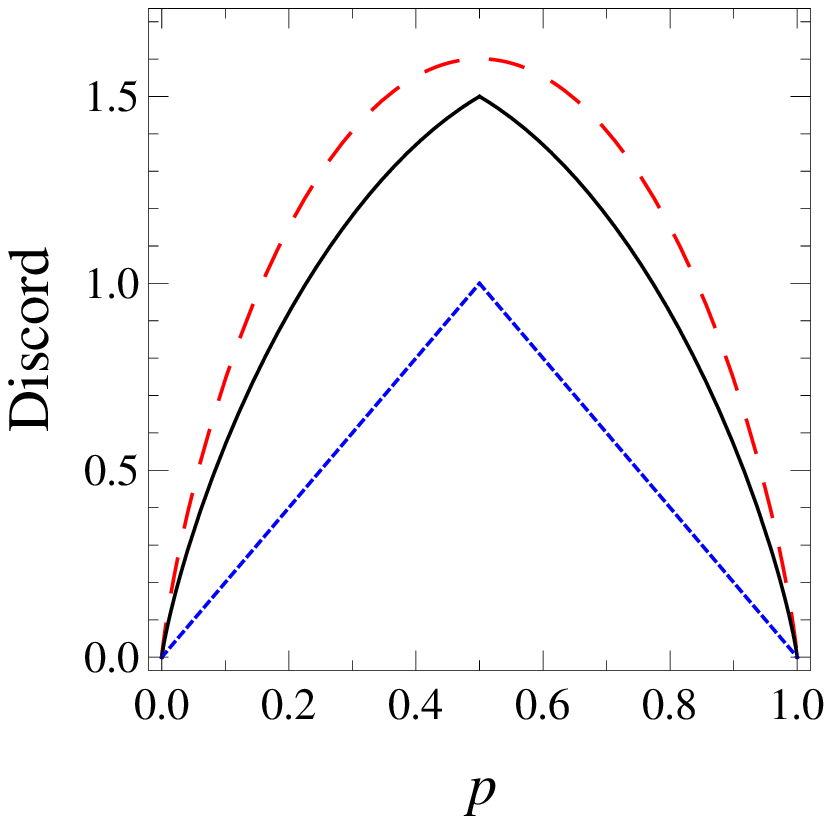}\label{S-b}}
\caption{(Color Online) Additivity relations of entanglement (left) and discord (right) for the states $\vert\phi\rangle=\sqrt{p}\vert 000\rangle +\sqrt{(1-p)}\vert \pm 11\rangle$, plotted versus $p$. Symbol ${\mathcal Q}$ is used to represent
$\mathcal{E}$ or $\mathcal{D}$.}
\end{figure}

While any bipartite subsystem of the GHZ state has zero discord, one may
apply an appropriate non-local unitary operation to the GHZ state and obtain
\begin{align*}
\frac{1}{\sqrt{2}}(\vert 000\rangle+\vert \mathcal{+}11\rangle)
\end{align*}
with nonzero discord $\mathcal{D}(A:B)$ and $\mathcal{D}(A:C)$.
Now we examine the additivity relations for a more general state $\vert \phi\rangle = \sqrt{p} \vert 000\rangle + \sqrt{1-p}\vert \mathcal{+} 11\rangle$. There is no entanglement in any bipartite subsystem of $\vert \phi\rangle$. We can obtain a representation for $\mathcal{E}(A:B:C)$ by using inequality \eqref{entanglement equality}. Let us consider a tripartite separable state $\sigma=p\vert000\rangle\langle000\vert+(1-p)\vert\mathcal{+}11\rangle\langle\mathcal{+}11\vert$, and it is clear that $\mathcal{E}(A:B:C)\leq S(|\phi\rangle\langle\phi|
\vert \vert \sigma)=-p\log p-(1-p)\log (1-p)$. However, it is also true that $\mathcal{E}(A:B)+\mathcal{E}(AB:C)=-p\log p-(1-p)\log (1-p)$, so $\mathcal{E}(A:B:C)=\mathcal{E}(A:B)+\mathcal{E}(AB:C)=-p\log p-(1-p)\log (1-p)$ from Eq.~\eqref{entanglement equality}. Bipartite entanglement is easily found as  $\mathcal{E}(AC:B)=-p\log p -(1-p)\log(1-p)$ and $\mathcal{E}(BC:A)=-\lambda_{+} \log \lambda_{+} -\lambda_{-}\log \lambda_{-}$ where $\lambda_{\pm}=1/2 \pm \sqrt{1/4-p(1-p)/2}$. Now one can check that all the additivity relations are satisfied. For the other permutations of $A$, $B$ and $C$, the results are the same except for a change of ``$=$'' in  \eqref{total equality} to ``$\geq$''. We plot $\mathcal{E}(A:B:C)$, $\mathcal{E}(BC:A)+\mathcal{E}(AC:B)$ and $\mathcal{E}(A:B)+\mathcal{E}(A:C)$ in Fig. \ref{S-a}.

In the case of the relative entropy of discord, one can show that $\mathcal{D}(A:B:C)= -p\log p-(1-p)\log(1-p)+\min\{p,(1-p)\}$ and $\mathcal{D}(A:B)=\mathcal{D}(A:C)=\min\{p,(1-p)\}$. Their proofs are straightforward (see Appendix). We also simply obtain $\mathcal{D}(BC:A)=-\lambda_{+} \log \lambda_{+} -\lambda_{-}\log \lambda_{-}$ and $\mathcal{D}(AC:B)=-p\log p -(1-p)\log(1-p)$, and find that the all the additivity relations are satisfied. For the other permutations, again, the same results are obtained except for the change from ``$=$'' to ``$\geq$'' in Eq.~\eqref{total equality}. The results are plotted in Fig. \ref{S-b}.
In addition, we examine the additivity relations for the measure of classical correlations. The classical correlations are obtained as
\begin{align*}
\mathcal{C}(A:B:C)=&-\frac{q}{2}\log\frac{q}{2}-(1-\frac{q}{2})\log(1-\frac{q}{2})\\&-q\log  q -(1-q)\log (1-q)-q,\\
\mathcal{C}(A:B)=&-\frac{q}{2}\log\frac{q}{2}-(1-\frac{q}{2})\log(1-\frac{q}{2})-q,\\
\mathcal{C}(AB:C)=&-q\log q -(1-q)\log (1-q),\\
\mathcal{C}(BC:A)=&-\lambda_{+} \log \lambda_{+} -\lambda_{-}\log \lambda_{-},
\end{align*}
,where $q=\min\{p,1-p\}$, and all the additivity relations are met.

\begin{figure}
\centering
\subfigure{\includegraphics[width=0.31\textwidth]{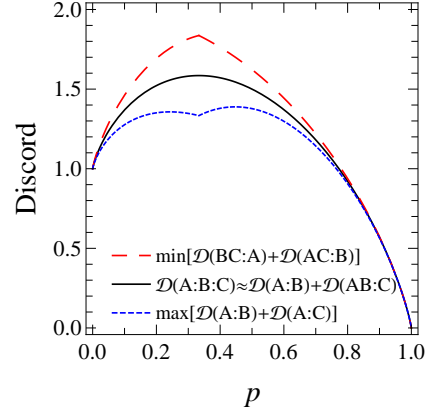}}
\caption{(Color Online) Additivity relations of discord for the generalized W-states $\sqrt{p} \vert 001\rangle +\sqrt\frac{1-p}{2}\big(\vert 010\rangle +\vert 100\rangle\big)$, plotted versus $p$. Here, the $\min$ ($\max$) stands for taking the minimum (maximum) over all permutations of A,B, and C.}\label{bounds1}
\end{figure}
We now consider a W-state
\begin{align*}
\vert W\rangle =\frac{1}{\sqrt{3}}(\vert001\rangle+\vert010\rangle+\vert100\rangle)
\end{align*}
with which it is known that $\mathcal{E}(BC:A)=\mathcal{E}(AC:B)=\log 3-2/3\approx 0.92$, $\mathcal{E}(A:B:C)=2\log3-2\approx 1.17$ \cite{Wei04}, and $\mathcal{E}(A:B)=\mathcal{E}(A:C)=\log 3 -4/3\approx 0.25$ \cite{Vedral98}. All the additivity relations are then satisfied.
In a more general case of
\begin{align*}
\alpha \vert 001\rangle +\beta\vert 010\rangle +\gamma\vert 100\rangle,
\end{align*}
the additivity relations of discord can be examined as follows. First, quantum discord with bipartition AB and C is easily obtained to $\mathcal{D}(AB:C)=-\vert\alpha\vert^{2}\log\vert\alpha\vert^{2}-(\vert\beta\vert^{2}+\vert\gamma\vert^{2})\log (\vert\beta\vert^{2}+\vert\gamma\vert^{2})$ because the total system is pure. $\mathcal{D}(A:C)$ and $\mathcal{D}(A:B:C)$ are upper bounded respectively by $-\vert\alpha\vert^{2}\log(\vert\alpha\vert^{2}/\vert\alpha\vert^{2}+\vert\gamma\vert^{2})-\vert\gamma\vert^{2}\log(\vert\gamma\vert^{2}/\vert\alpha\vert^{2} +\vert\gamma\vert^{2})$ and $-\vert \alpha \vert ^{2}\log\vert \alpha \vert ^{2}-\vert \beta \vert ^{2}\log\vert \beta \vert ^{2}-\vert \gamma \vert ^{2}\log\vert \gamma \vert ^{2}$ as we choose the logical basis in the expression \eqref{quantumness 2}. The inequality \eqref{total bound 1} is then satisfied for discord as
 \begin{align*}
  \mathcal{D}(AB:C)-\mathcal{D}(A:C) \geq &-(\vert\alpha\vert^{2}+\vert\gamma\vert^{2})\log(\vert\alpha\vert^{2}+\vert\gamma\vert^{2})\\ &-(\vert\beta\vert^{2}+\vert\gamma\vert^{2})\log(\vert\beta\vert^{2}+\vert\gamma\vert^{2})\\&+\vert\gamma\vert^{2}\log \vert\gamma\vert^{2} \geq 0.
  \end{align*}
  The last inequality holds because for given $\vert\gamma\vert$, $-(\vert\alpha\vert^{2}+\vert\gamma\vert^{2})\log(\vert\alpha\vert^{2}+\vert\gamma\vert^{2})  -(\vert\beta\vert^{2}+\vert\gamma\vert^{2})\log(\vert\beta\vert^{2}+\vert\gamma\vert^{2})$ is minimized at $\alpha=0$ or $\beta=0$. The inequality \eqref{total bound 3} is similarly satisfied as
   \begin{align*}
  \mathcal{D}(A&B:C)+\mathcal{D}(A:BC)-\mathcal{D}(A:B:C)\\ \geq &-(\vert\alpha\vert^{2}+\vert\beta\vert^{2})\log(\vert\alpha\vert^{2}+\vert\beta\vert^{2})\\ &-(\vert\beta\vert^{2}+\vert\gamma\vert^{2})\log(\vert\beta\vert^{2}+\vert\gamma\vert^{2})\\&+\vert\beta\vert^{2}\log \vert\beta\vert^{2} \geq 0.
  \end{align*}
  The inequalities also hold for the other permutations of A-B-C, because the procedure is invariant under permutations.
  In Fig. \ref{bounds1}, we plot the numerical result for the special case $\vert\alpha\vert^{2}=p$ and $\vert\beta\vert^{2}=\vert\gamma\vert^{2}=\frac{1-p}{2}$. It shows that $\mathcal{D}(A:B:C)\approx\mathcal{D}(A:B)+\mathcal{D}(AB:C)$, so Eq. \eqref{total equality} approximately holds .

\subsection{Additivity relations of discord in mixed states}
\begin{figure}
\centering
\includegraphics[width=0.31\textwidth]{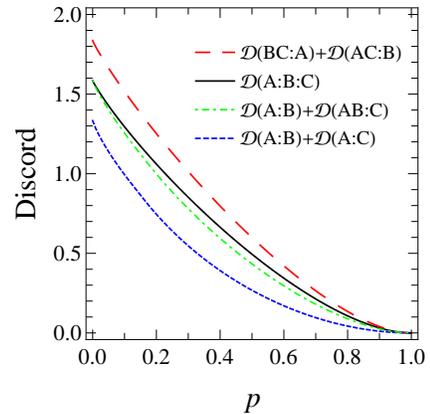}
\caption{(Color Online) Additivity relations of discord for the mixed state $\rho=(1-p)\vert W\rangle \langle W\vert + p \mathcal{I}/8$, plotted versus $p$.}\label{wmixture}
\end{figure}

   Our arguments about the additivity relations of discord were restricted to pure states. We noted that the additivity relations of discord are violated by some mixed systems whose ordering does not satisfy $\mathcal{D}(A:B)\geq\mathcal{D}(B:C)\geq\mathcal{D}(A:C)$. In this subsection, we examine a couple of mixed state examples states in order to check whether the additivity relations of discord hold.
   For the state
   \begin{align*}
   \rho=(1-p)\vert GHZ\rangle \langle GHZ\vert +p\mathcal{I}/8
   \end{align*}
   where $\mathcal{I}$ is the identity matrix, $\mathcal{D}(A:B:C)=\mathcal{D}(BC:A)$ and $\mathcal{D}(A:B)=0$, so the additivity relations are trivially met. The state
   \begin{align*}
   \rho=(1-p)\vert W\rangle \langle W\vert +p\mathcal{I}/8
   \end{align*}
   also satisfies the additivity relations and it is shown in Fig. \ref{wmixture}. The state $\rho=(1-p)\vert W\rangle \langle W\vert +\frac{p}{2}(\vert 000\rangle\langle000\vert +\vert 1\mathcal{+}1\rangle \langle 1\mathcal{+}1\vert)$
    satisfies the ordering required, so Eq.~\eqref{total bound 2} is satisfied for all $p$, but $\mathcal{D}(A:B:C)\ngeq\mathcal{D}(A:B)+\mathcal{D}(B:C)$ for $p> p^{*}$ where $p^{*}\approx 0.8$.

\section{Conclusion}
  We have studied about four additivity relations of the measures of multipartite correlations which are based on the quantum relative entropy. The relations were motivated by the definition of the total mutual information and the strong subadditivity of von Neumann entropy. They specify multipartite correlations in terms of bipartite correlations in subsystems, so they give insight of to what extent multipartite correlations contain bipartite correlations of subsystems. We have identified the additivity relations of entanglement, and argued that discord has the additivity relations for pure states, but it is not always true for mixed states.

\section{Acknowledgements}
 This work was supported by the National Research Foundation of Korea (NRF) grant funded by the Korean Government (No. 3348-20100018).
\section{Appendix}
 Let us compute the relative entropy of discord of the state $\rho=p\vert000\rangle\langle000\vert+(1-p)\vert1\mathcal{+}1\rangle\langle1\mathcal{+}1\vert$. First, we argue that taking minimization over all local base of B is enough to obtain $\mathcal{D}(A:B:C)$ of the state. Consider a measure of discord $\mathcal{D}^{*}(\rho_{ABC})=\min _{\sigma\in\mathcal{C}^{*}}S(\rho\vert\vert\sigma)$ where $\mathcal{C}^{*}$ is the set of all quantum states in the form $\sigma=\sum p_{i}\vert i\rangle_{B}\langle i\vert\otimes\sigma_{AC}^{i}$ where $\{\vert i\rangle_{B}\}$ is a local basis of B, and $\sigma_{AC}^{i}$'s are arbitrary quantum states of AC. Apparently, $\mathcal{D}^{*}\leq\mathcal{D}$. It can be shown that the minimum is obtained for a state of the form $\sigma=\sum \vert i\rangle \langle i \vert \otimes \langle i\vert\rho\vert i \rangle$, as $\mathcal{D}$ can be reduced to the expression \eqref{quantumness 2}. On the other hand, in the case of our target state, $\sigma$ is also a classically correlated state for every choosing of the local basis of B. Therefore $D^{*}=D$ in this case. Now, local base of B can be parameterized as Eq.~\eqref{parameterization}, and simple differentiations find two local minima of $S(\rho\vert\vert\sigma)$ at $\sigma^{*}=p\vert000\rangle\langle000\vert+\frac{(1-p)}{2}(\vert101\rangle\langle101\vert+\vert111\rangle\langle111\vert)$ and $\sigma^{**}=\frac{p}{2}(\vert0\mathcal{+}0\rangle\langle0\mathcal{+}0\vert)+\vert0\mathcal{-}0\rangle\langle0\mathcal{-}0\vert)+(1-p)\vert1\mathcal{+}1\rangle\langle1\mathcal{+}1\vert$. Consequently, $\mathcal{D}(A:B:C)=\mathcal{D}(A:B)=\min\{p,(1-p)\}$, and then discord of the state $\vert\psi\rangle=\sqrt{p}\vert000\rangle+\sqrt{1-p}\vert1\mathcal{+}1\rangle$ is computed to $\mathcal{D}(A:B:C)=-p\log p-(1-p)\log(1-p)+\min\{p,(1-p)\}$ by applying \eqref{discord equality}.


\begin{thebibliography}{1}
\bibitem{footnote1} $p_i$ can be a probability of an arbitrary system. If it is a probability of system A-B-C, it can be denoted as $p_{ABC}(i,j,k)$. Once the probabilities of the total system are given, the probabilities of the subsystem can be given as its marginal as, e.g., $p_{AB}(i,j)=\sum_k p_{ABC}(i,j,k)$, $p_{AC}(i,k)=\sum_j p_{ABC}(i,j,k)$, $p_{BC}(j,k)=\sum_i p_{ABC}(i,j,k)$, and so on.
\end{thebibliography}
\end{document}